# Size induced metal insulator transition in nanostructured Niobium thin films: Intragranular and intergranular contributions

Sangita Bose,[1] Rajarshi Banerjee,[2] Arda Genc,[3] Pratap Raychaudhuri,[1] Hamish L. Fraser[3] and Pushan Ayyub[1]

[1]Department of Condensed Matter Physics and Material Science, Tata Institute of Fundamental Research, Mumbai 400005, India.

[2]Department of Materials Science and Engineering, University of North Texas, Denton, Texas 76203, USA

[3]Department of Materials Science and Engineering, Ohio State University, Columbus, Ohio 43210, USA

**Abstract**: With a reduction in the average grain size in nanostructured films of elemental Nb, we observe a systematic crossover from metallic to weakly-insulating behavior. An analysis of the temperature dependence of the resistivity in the insulating phase clearly indicates the existence of two distinct activation energies corresponding to inter-granular and intra-granular mechanisms of transport. While the high temperature behavior is dominated by grain boundary scattering of the conduction electrons, the effect of discretization of energy levels due to quantum confinement shows up at low temperatures. We show that the energy barrier at the grain boundary is proportional to the width of the largely disordered inter-granular region, which increases with a decrease in the grain size. For a metal-insulator transition to occur in nano-Nb due to the opening up of an energy gap at the grain boundary, the critical grain size is $\approx$ 8nm and the corresponding grain boundary width is $\approx$ 1.1nm.

## 1. Introduction

A majority of the novel physico-chemical properties exhibited by nanomaterials [1,2] is attributed to either surface effects or quantum confinement effects (size-dependent changes in the electronic band structure) [3-7] Several groups have made detailed investigations of the electrical and structural properties of nanomaterials. Systematic studies of nanocrystalline metals and alloys such as Cu, Pd, Fe, Pt [1], Ag [8] and NiP [9] usually show an increase in both the room temperature resistivity and the residual resistivity ($T\rightarrow0$) with decreasing size, and the resistivity is roughly proportional to the absolute temperature over a wide temperature range. The temperature coefficient of resistance (TCR) is also found to decrease with a reduction in size. This is generally explained in terms of an enhanced scattering from the grain boundaries, which are modeled as potential barriers of a certain height and width [1]. Electron scattering theory predicts a linear relation between electrical conductivity and grain size. Though a negative temperature coefficient of resistance has been predicted, it has been actually observed in very few nanocrystalline metallic systems [10].

The size dependence of superconducting properties has been investigated experimentally in many zero dimensional systems such as Sn [11], Al, Pb and In [7,12]. Weak coupling type I superconductors such as Al and Sn show an increase in the superconductor transition temperature ($T_C$), which is thought to be caused by a softening of surface phonon modes [13]. However, a strong coupling, type-I superconductor such as Pb does not show any change in $T_C$ down to a particle size of ~6nm [14] It appears that both surface effects and the quantum size effects (QSE) independently affect the superconducting properties of low dimensional systems. From detailed spectroscopic measurements on single Al nanoparticles, Ralph *et al*. [15] established that the superconducting order parameter can indeed exist for sizes much below the coherence length and vanishes only when the discrete energy level spacing exceeds the superconducting energy gap. However, the effect of the discretization of levels on the superconducting properties such as the transition temperature and superconducting gap remains to be explored in detail.

Structurally, nanostructured materials can be visualized as nanometer-sized grains separated by grain boundaries. The structure of the grain boundaries in nanocrystalline materials has invoked considerable controversy in the past decade [16] Some workers [17,18] claim that nanocrystalline grain boundaries can be almost atomically sharp and essentially resemble those in bulk polycrystalline materials, while others suggest that they are thicker and exhibit a 'gas-like' structure, virtually devoid of even short-range order [19-21]. A recent, detailed study based on extended x-ray absorption fine structure shows that the grain boundary structure in nanocrystalline ZnO films is neither similar to bulk polycrystalline materials nor gas-like [22] The atomic arrangement in these grain boundaries has short-range order extending to the second coordination shell, but no long-range order. In this context, one must recognize that the actual grain boundary structure in a given nanostructured material would be substantially influenced by the processing method.

In a recent communication, we have shown that the size dependence of superconductivity in nanostructured Nb is primarily governed by the changes in the electronic density of states and not the electron-phonon coupling due to surface effects [23]. In the present paper we report the size-dependant evolution of the normal state transport properties and the superconducting transition in nanostructured Nb films and show how these properties are influenced by the microstructure of the granular films. From x-ray diffraction (XRD), high-resolution transmission electron microscopy (HRTEM) and electrical transport measurements, we can identify three distinct size regimes. Nanocrystalline films with an average grain size greater than 28nm behave essentially like bulk metals, exhibiting the bulk superconducting transition temperature ($T_C$) of 9.4K for Nb. These films consist of closely packed grains that have strong inter-granular coupling. Films with grain size in the 8-28nm range show a metallic behavior but there is a size-dependent decrease in the $T_C$ from 9.2K to 4.7K. In this regime, the grains are separated by partially insulating intergranular regions, which act as potential barriers separating the individual grains. Finally, in grains smaller than 8nm, we observe a weakly insulating behavior inferred from the activated nature of the temperature dependent resistivity. The grains in these films have a rather extensive intergranular region (> 1nm) as evidenced from HRTEM. Furthermore, in this regime, we clearly observe *two* distinct activation energies associated with intergranular and intragranular transport in nanostructured Nb. We thus show that the transport properties of nanostructured Nb are governed not only by quantum size effects, but are also strongly influenced by the nature and extent of the grain boundary region, which have to be taken into account to understand the evolution of transport properties with decreasing particle size.

**2. Experimental details**

Nanostructured Nb thin films were deposited on Si substrates (with an amorphous oxide overlayer) using high pressure dc magnetron sputtering [24]. Sputter deposition was carried out from a 99.99% pure Nb target in a custom-built UHV chamber with a base pressure $\sim 3\times 10^{-8}$ torr. The average grain size in the nanostructured films could be controlled by varying the sputtering gas (Ar) pressure in the 3-100 mtorr range and the dc power in the 25-200W range [25]. The Ar gas was passed through a getter furnace with Ti sponge maintained at 1073K prior to introduction in the chamber. The crystallographic structure and the microstructure of the as-deposited nano-Nb films were investigated by x-ray diffraction (XRD) and transmission electron microscopy (TEM). We used a Siemens D500 x-ray diffractometer in the θ-2θ geometry with CuK$_\alpha$ radiation, and a FEI/Philips CM200 TEM and a FEI Tecnai F20 FEG-TEM, both operating at 200 kV. We carried out an x-ray line profile analysis using the X-FIT software, which uses the Warren-Averbach method to obtain the strain corrected coherently diffracting domain size ($d_{XRD}$). By controlling the sputtering parameters (see above), $d_{XRD}$ could be varied from 60 nm down to 5 nm. The thickness of the nanostructured films was measured from cross-sectional TEM and was found to vary between 0.5μm and 0.8μm. The temperature dependence of the resistance was measured using the four-probe technique in a liquid-He dewar, in which the temperature could be varied from 4.2 to 300K. The superconducting $T_C$ was obtained from measurements of both dc magnetization (Meissner effect) and electrical transport. Magnetization measurements were carried out in a MPMS Quantum Design SQUID magnetometer.

## 3. Results

*3.1 Structural properties*

Table I lists the combination of sputtering parameters employed to synthesize the nano-Nb thin films with average crystallographic domain size ($d_{XRD}$) in the 5-60 nm range. No impurity phases (e.g., oxides) were seen in the XRD spectra of any of the films down to the smallest grain size. Figure 1 shows the [110] diffraction line corresponding to the *bcc* phase of Nb, and no other phase was seen in any of the samples. This was corroborated by electron diffraction data, which also showed the presence of only the *bcc* phase of Nb [25]. As expected, Fig. 1 shows an increase in the full width at half maximum of the [110] line with decreasing grain size. The upper limit of the particle size distribution obtained by the Warren-Averbach method (x-ray line profile analysis) was found to be no more than 20% of the mean width for any of the samples. From Fig. 1, we also observe a pronounced shift in the [110] line to lower values of 2θ with decreasing $d_{XRD}$, indicating a size-induced expansion in the unit cell in nano-Nb. The monotonic increase of the cubic lattice constant in nano-Nb with decreasing grain size was reported by us in an earlier communication [25] and discussed in detail in terms of a linear elasticity model.

Figures 2(a) and 2(c) show, respectively, the bright-field TEM micrographs of the nano-Nb thin films with the largest (≈ 60nm) and the smallest (≈ 5nm) grain sizes in our series. The diffracting grains exhibiting the *darkest* contrast indicate the typical grain sizes and shapes. Figures 2(b) and 2(d) show the corresponding dark field TEM images of the same samples.

Since these images show only grains with a particular crystallographic orientation, it is easier to identify individual grains. While the grain sizes observed in the dark-field images of all the samples typically showed a close agreement with the calculated values of $d_{XRD}$, a few grains in the dark-field image (see, e.g., Fig. 2(d)) appeared to be much larger, possibly due to the overlap of bright contrast arising from multiple diffracting grains of similar crystallographic orientation. A comparison of Figures 2(a) and 2(c) clearly indicates that the larger grains have sharper interfaces while the smaller grains are separated by distinct inter-granular regions. This description is particularly true of the samples with smaller average grain sizes. In the HRTEM images of the sample with $d_{XRD}$ = 5nm (Figure 3), we can identify a couple grains in which crystalline order persists for ~ 5-6 nm. The inter-granular region is identified by an absence of long-range order, but with some degree of short-range order. The width of the inter-granular region is ~ 1nm in this film. This type of grain boundary is structurally similar to that recently observed in ZnO films [22].

The nature of the inter-granular region (IGR) was probed in further detail using electron energy loss spectroscopy (EELS). Figure 4 shows the EELS spectra recorded from the following regions: (i) interior of a 18 nm grain (top curve), (ii) IGR of 18 nm grain (second from top), (iii) interior of a 5 nm grain (third), and (iv) IGR of 5 nm grain (bottom). The inset shows the standard EELS spectrum for $Nb_2O_5$ from the EELS atlas. The EELS spectra shown here correspond to the low energy loss region (< 100 eV). The first strong peak in each spectrum is the zero loss peak (from electrons that do not interact with the specimen). The second and third peaks are plasmon peaks due to electrons that interact with the free electrons in the material, and indicate the free-electron density in the system. It is difficult to do a quantitative analysis due to the presence of two plasmon peaks in the low-loss spectra, but a qualitative analysis is possible. The EELS spectra from the grain interiors of both the 18 nm and 5 nm grains are similar and appear substantially different from the $Nb_2O_5$ standard spectrum. In contrast, the spectra for the IGR of both samples are quite similar to that of $Nb_2O_5$ (see inset). Thus the inter-granular regions in the nano-Nb samples appears to be made up of an amorphous phase rich in niobium and oxygen, possibly in the form of an amorphous oxide. We can therefore describe the microstructure of the nano-Nb system in terms of isolated crystalline grains of *bcc* Nb in a structurally disordered matrix. In the next section, we show that the in films with $d_{XRD} < 8$nm, such grain boundaries can act as insulating barriers giving rise to an activated transport behavior.

*3.2 Electrical properties*

Figure 5(a) shows the temperature dependence of the resistivity in nanostructured Nb films with different grain sizes. While the room temperature resistivity, $\rho_{300} = 7\mu\Omega$cm, for the sample with $d_{XRD} = 60$nm is comparable to that of bulk Nb, there is an almost three orders of magnitude increase in $\rho_{300}$ to 14m$\Omega$cm for the sample with $d_{XRD} = 5$nm. Based on their transport properties, we can categorize the nanostructured Nb films into three different size regimes. Nano-Nb films with $d_{XRD} > 28$nm behave like bulk, metallic Nb with a positive TCR between $T_C$ and 300K. Samples with 28nm $\geq d_{XRD} \geq$ 11nm also show a metallic behavior with a positive TCR. However, in this regime, the grain boundary acts as an insulating barrier, as observed from the presence of hysteresis in the critical current in the superconducting state. Finally, films with grains $d_{XRD} < 8$nm show a weakly insulating behavior with negative TCR, while the film with $d_{XRD}$ = 8nm shows an almost temperature independent resistivity. We now discuss the transport properties in each of these regimes separately.

*1. Regime I ($d_{XRD}$ > 28nm)*

In this regime the films display essentially bulk properties as far as the normal state resistivity and the TCR are concerned. Nano-Nb samples in this size range superconduct with the same $T_C$ as in bulk Nb (9.4K). These samples also show the bulk coherence length of 41nm, as obtained from magnetization measurements, details of which would be presented in a separate communication [26]. The TEM images of these samples (Fig 2(a) and (b)) show that the crystalline Nb grains are separated by sharp interfaces and hence also structurally resemble a bulk polycrystalline film.

*2. Regime II (28nm $\geq d_{XRD} \geq$ 11nm)*

Nanostructured Nb thin films in this size regime continue to superconduct, but there is a monotonic, size-induced decrease in the $T_C$ from 9.2K to 5.9K. TEM and EELS data indicate that the nanocrystalline grains in this regime are isolated and separated by thin intergranular regions of an amorphous phase containing niobium and oxygen. From the perspective of electron transport, therefore, these films can be considered as ensembles of metallic nano-grains embedded in a poorly conducting disordered matrix. The roughly six-fold increase that is observed in the value of $\rho_{300}$ when the size is decreased from 28nm to 11nm (see Fig. 5(b)), can be explained by the following intuitive argument. A reduction in the particle size by a factor of three would lead to (i) a three-fold increase in the number of grain boundaries (per unit volume) that are expected to act as scattering centers, and (ii) a three-fold decrease in the mean free path, provided it is limited by the grain size alone. This should lead to a nine-fold increase in the resistivity. However, if the mean free path is not purely grain-size limited and decreases less slowly than the grain size, we would expect to see a somewhat smaller increase in the room temperature resistivity. Thus, the six-fold increase in $\rho_{300}$ appears reasonable.

In an attempt to further validate our claim that the nanostructured films in this size regime behave as a granular metal, we carried out an *I-V* measurement using the standard four-probe method at temperatures below their respective $T_C$'s. The I-V characteristics of the films in this size range exhibited a hystersis, [23] which implies, according to the RCSJ model, [27] that the films form a random array of weakly coupled Josephson junctions. Similar behavior has been observed earlier in quench-condensed films of Pb and Bi [28]. Expectedly, no such hystersis behavior was seen in the I-V characteristics of the film with $d_{XRD} \geq 28$nm, which behaves as a bulk crystalline material with closely packed grains. The intergranular boundaries in the latter films are transparent and hence metallic in the normal state. Besides, the resistance in these films drops to zero below the $T_C$ (note that the $T_C$ obtained from magnetization and resistivity measurements coincide), showing that the inter-granular phase does not form a percolative network. Rather, the system consists of a collection of individual grains, within which the electronic wave functions are confined.

The monotonic decrease in the $T_C$ in this size regime can be understood from the fact that there is a clear correlation between $T_C$ and the superconducting energy gap, $\Delta(0)$, which we have directly measured by point contact spectroscopy.[23] Nanocrystalline Nb continues to behave as an intermediate-coupling, type II superconductor down to the lowest size of 11nm with $2\Delta(0)/k_BT_C \sim 3.6$. From this we infer that the changes in $T_C$ are governed by the changes in the electronic density of states arising in nano grains from the quantization of the electronic wave vector. This is consistent with the theoretical calculations of Strongin *et al.* [29] who have invoked the discretization of the energy levels in the BCS equation and predicted a decrease in $T_C$ in dimensionally-confined superconductors.

*3. Regime III ($d_{XRD}$ < 8nm)*

Nano-Nb samples with $d_{XRD}$ < 8nm show a negative temperature coefficient of resistance and a weak, activated behavior. These samples are also non-superconducting, consistent with the Anderson criterion [30]. Anderson had predicted in 1959 that superconducting order is destabilized in grains that are so small that the level spacing between the discretized energy bands (known as the Kubo gap, δ) is comparable to the superconducting energy gap. For Nb, this critical size can be calculated to be 7nm, which almost exactly agrees with our experimental observation (see Fig. 5(a)).

We now attempt to analyze and understand the metal-insulator transition observed in the electrical transport data (Fig. 5). Considering quantum size effects (QSE) alone, we should expect a metal-insulator transition to occur when the Kubo gap, δ ~ $4\varepsilon_F/3N$, (N = number of conduction electrons at the Fermi energy, $\varepsilon_F$) becomes comparable to $k_BT$ [3] The QSE-driven metal-insulator transition is expected at ~35K for 5nm sized Nb nanoparticles, and at ~15K for 7nm particles. However, we observe the onset of the insulating behavior at a much higher temperature in samples with $d_{XRD}$ < 8nm. Qin *et al.* [8] have also seen a negative TCR in nano-Ag for grain sizes < 9nm and a density of ~ 45-50%, but attribute their observation to a high density of microscopic vacancy-like defects at the grain boundaries.

## 4. Discussions

To understand the origin of the negative TCR and the high metal-insulator transition temperature in nano-Nb, we attempted to fit the σ-T curves ($d_{XRD} \leq$ 7nm) with an activated transport behavior. We used the following three types of empirical trial functions: (i) $\sigma = A\exp(-E_g/k_BT)$, (ii) $\sigma = \sigma_0 + A\exp(-E_g/k_BT)$ and (iii) $\sigma = \sigma_0 + A\exp(-E_{g1}/k_BT) + B\exp(-E_{g2}/k_BT)$, where $\sigma = \rho^{-1}$. The activation energies ($E_{gi}$) and the proportionality constants A and B are used as best fit parameters. The first one, which corresponds to a simple activation barrier with a single gap, clearly does not fit our data. The best fit of equation (ii), which contains a temperature-independent contribution ($\sigma_0$) in addition to the activated behavior, to the σ-T data for the 5nm Nb sample is shown by the dashed curve in Fig. 6(a). The corresponding deviation from the data is shown - also by the dashed curve - in Fig. 6(b). Obviously, the fit is far from satisfactory, particularly in the lower temperature region. However, an attempt to fit the data to equation (iii) which contains two activation energies, $E_{g1}$ and $E_{g2}$ (in addition to the temperature-independent contribution), produced much more satisfactory results, as shown by the solid curve in Fig. 6(a). A comparison of the deviation of the σ vs $1/k_BT$ data points from the single and double exponential fits for the 5nm sample (Fig 6(b)) clearly indicates that the fit to the double exponential is significantly better, particularly at low temperatures. Figure 6(c) shows the double exponential fits to the transport data for all the three insulating samples. The fit is good down to the lowest temperature (4.2K). The high temperature downturn in samples with $d_{XRD}$ = 6 and 7nm comes from the temperature region in which the samples exhibit metallic behavior. We now show that the double exponential character of the conductivity has a convincing physical basis.

What is the physical origin of the two activation energies ($E_{g1}$ and $E_{g2}$) connected to the transport behavior of the insulating phase of nanostructured Nb? The values of $\sigma_0$, $E_{g1}$ and $E_{g2}$ obtained from the double exponential fits and the calculated Kubo gaps ($\delta$) for the three samples with the smallest particle sizes are listed in Table 2. Both the characteristic activation energies increase monotonically with decreasing particle size. From a comparison of their respective magnitudes and size dependences, we identify the lower of the two energy gaps ($E_{g1}$) with the Kubo gap arising from the size-induced discretization of the energy levels (the small discrepancy can probably be attributed to electron correlation effects). A physical estimation of this gap is important as it provides the energy scale for the validity of the Anderson criterion for low-dimensional superconductors. However, the manifestation of the Kubo gap would only be appreciable at low temperatures, and is certainly not responsible for the insulating behavior close to room temperature.

While we have argued that the smaller of the two activation energies has a purely intra-granular quantum mechanical origin, the larger one appears to be related to inter-granular electronic transport. This can be clearly seen by comparing the characteristic temperatures $T_i = E_{g2}/k_B$ obtained from $E_{g2}$ for the three insulating samples (see Table 2) with the inflexion temperatures in the resistivity data. The higher of the two activation energies ($E_{g2}$) is therefore associated with the metal-insulator transition in the smaller-sized nanostructured samples ($d_{XRD} < 8$nm). We now show that $E_{g2}$ (whose value lies in the 16–31 meV range) appears to originate from the potential barrier at the grain boundaries that needs to be overcome for inter-grain conduction to occur. Since the 5nm grain has a grain boundary width, $\Delta d > 1$nm (see the HRTEM images in Fig. 3), while large-grained samples are known to have atomically sharp boundaries, it is clear that $\Delta d$ is inversely related to the grain size in this system. This results in a *size-dependent* potential barrier that leads to the activated conduction at room temperature. Such a model also agrees with the explanation given by Qin *et al.* [8] for the negative TCR in nanocrystalline Ag. A larger $\Delta d$ implies an increase in the vacancy-like defects that tend to make the sample comparatively insulating.

Due to some degree of variability in the grain size as well as the grain boundary width in the same sample, the relation between the grain size and the grain boundary width ($\Delta d$) cannot be accurately determined solely from the TEM images. We therefore estimated the grain boundary width using the following procedure. We have earlier shown [25] that the cubic lattice constant in nano-Nb increases monotonically with decreasing grain size. Using an equation derived from linear elasticity theory, the lattice expansion can be related to the grain size, in terms of $\Delta d$ and other parameters. This relation allows one to obtain $\Delta d$ as a function of grain size (Fig. 7). Significantly, we observe a linear relation between the upper barrier energy ($E_{g2}$) and $\Delta d$ (Fig. 7, inset), an extrapolation of which suggests that the value of $\Delta d$ corresponding to *zero* barrier energy is $\approx 1.0$nm. (Note that 'zero barrier energy' corresponds to relatively large sized grains with sharp grain boundaries and, consequently, metallic properties.) The grain size corresponding to this value ($\approx$ 1nm) $\Delta d$ is $\approx$ 9nm, in excellent agreement with the *experimentally observed* critical particle size at which the metal-insulator transition is seen to occur (8nm). This is also evident from Fig 5(b), which shows a sharp increase in the resistivity below 8nm.

Assuming the grain boundary structure to be largely amorphous with an ordered first coordination shell, a partially ordered second coordination shell, and disordered higher shells (as per recent EXAFS data [22]), we would expect the energy gap at the grain boundary to open up for a boundary width ≈ second coordination shell. The interatomic distance in Nb is 0.286 nm, and therefore the second and third coordination shells correspond to a spacing of ≈ 0.57 nm and ≈ 0.86 nm from the outermost fully ordered layer of the grain. The grain boundary is expected to be partially ordered if it is limited to the second coordination shells of adjacent grains leading to a boundary width of ≈ 1.14 nm. If the boundary width is larger, we can expect the inter-granular region to be structurally disordered. Thus, an energy gap in the electrical transport should open when $\Delta d \geq 1.14$ nm, which is in very good agreement with the experimentally determined, limiting value of $\Delta d$ of about 1 nm.

Note also that as $T \rightarrow 0$, $\sigma \rightarrow \sigma_0$, a constant value. A physical interpretation of the temperature-independent $\sigma_0$ is less straightforward than $E_{g1}$ and $E_{g2}$. The finite conductivity at zero temperature suggests the existence of an additional transport channel for which the electrons do not experience either of the above two activation barriers. This could originate from two sources: (i) The irregular shape and the size distribution of the particles size is quite likely to introduce a significant number of mid-gap states within the Kubo gap, and (ii) we also may have a distribution of grain boundary widths, with a small fraction of the grain boundaries which are much more strongly connected than the rest. An electron passing across the strongly connected grain boundaries would essentially undergo metallic transport and not experience the potential barrier $E_{g2}$. These electrons would therefore give rise to a finite conductivity even at the lowest temperatures. In this context, we point out that the resistivity of structurally disordered (amorphous) metals and alloys remains finite as $T \to 0$.

Transport in granular systems such as Au/SiO$_2$, W/Al$_2$O$_3$, Ni/SiO$_2$ [31], in which there are metallic nanograins (< 4nm) in a large insulating matrix, has been studied for a long time. In many of these systems, the resistivity data have been fitted to the relation $\sigma = \sigma_0 \exp\left[-(T/T_0)^{-1/2}\right]$ obtained from Mott's variable range-hopping (VRH) model, taking electron correlations into account. However, our data cannot be fitted to the (VRH) model (with or without electron correlations). To show this, we have plotted the log of the conductivity with $T^{-1/4}$ and $T^{-1/2}$ as shown in Figures 8(a) and 8(b) respectively. The fitting clearly shows that the conductivity is proportional to neither $T^{-1/4}$ nor $T^{-1/2}$ over the entire temperature range of 4.2-300K. According to a recent study, freshly prepared nanostructured Cu does show VRH behavior, while nanoparticles with appreciably oxidized grain boundaries show an activated behavior [32]. Since the inter-granular region in our films is known to be partially oxidized, we not observe VRH-type behavior. We also point out that many previous reports on the metal-insulator transition showing a $T^{-1/2}$ dependence of the conductivity actually dealt with two-dimensional films (usually quench condensed) [33], in which the surface resistance dictates the transition. Our films are relatively thick (~0.5μm) and consist of stacks of weakly connected nanoparticles. These are essentially three-dimensional nanostructures, in which the volume resistivity (and not the surface resistance) is the relevant parameter.

It is therefore clear that the existing models based on variable range hopping conduction, that have been used to explain transport in granular or nanostructured systems do not fit our data. The model proposed by us is based on a size-dependent grain boundary width and not only describes our experimental results adequately, but is also quantitatively consistent with our data on size-dependent lattice expansion. The TEM data also supports our notion of smaller grain-sized samples having wider grain boundaries.

## 5. Conclusions

The electrical conductivity in sputter-deposited nanostructered Nb films shows an interesting dependence on the crystallographic grain size and the microstructure. For grain sizes above 28 nm, Nb behaves as a bulk superconductor with a normal metallic phase (i.e., a positive temperature coefficient of resistance). Between 28nm and 11nm, the superconducting $T_C$ decreases and the normal state resistivity increases with a reduction in size. Below 8nm, the system no longer superconducts and exhibits a size-dependent metal-insulator transition, with a change in the sign of the TCR. In this size range, the conduction electrons in nanostructured Nb are weakly localized within the grains due to structurally disordered intergrain boundaries. The grain boundary width and consequently the height of the energy barrier corresponding to the grain boundaries are size dependent. Such a situation leads to a size-induced transition from a metallic to a weakly-insulating phase.

Our transport data clearly indicate that the conductivity in the smallest temperature regime is associated with two distinct activation energies, originating from quite different physical phenomena. While the larger activation energy is related to hopping across poorly conducting grain boundaries, the smaller one originates from the Kubo gap due to discretization of the energy levels in small grains. The effect of the smaller gap is also manifested in the observed disappearance of superconductivity, consistent with Anderson's prediction. The critical grain size ($d_{XRD}$) at which an insulating energy gap opens up is $\approx$ 8nm and corresponding grain boundary width is $\approx$ 1.1nm. These values, obtained from our model, match our observations very well.

**Acknowledgement**

SB would like to thank the TIFR Endowment Fund for partial financial support.

**FIGURE CAPTIONS**

**Figure 1.** XRD pattern showing [110] reflection of *bcc* Nb for nanostructured Nb films with different values of coherently diffracting domain size ($d_{XRD}$). (Color online)

**Figure 2.** (a) Bright field and (b) dark field TEM images of the nanostructured Nb film with $d_{XRD} \approx 60$nm. (c) Bright field and (d) dark field TEM images of the film with $d_{XRD} \approx 5$ nm.

**Figure 3.** High-resolution transmission electron micrographs of the nanostructured Nb sample with $d_{XRD} \approx 5$nm. The grains are defined as crystallographically ordered regions (delineated in the figure) while the disordered inter-granular regions are $\approx 1$nm wide.

**Figure 4.** Electron energy loss spectra recorded (top to bottom, respectively) from the following specific regions i) interior of grains with $d_{XRD} = 18$nm, ii) intergranular region of the 18nm sample, iii) interior of grains with $d_{XRD} = 5$nm, and iv) intergranular region of the 5nm sample. A standard EELS spectrum from $Nb_2O_5$ is shown in the inset.

**Figure 5.** (a) Temperature dependence of the resistivity for nanostructured Nb films with different grain sizes ($d_{XRD}$). The scale on the left refers to the metallic films with $d_{XRD} \geq 8$nm while the scale to the right is for the insulating films with $d_{XRD} < 8$nm. (b) Particle size dependence of the resistivity at 10K and 300K, depicted, respectively by open triangles and circles. (Color online)

**Figure 6.** (a) Comparison of single (dashed line) and double (solid line) exponential fits to the resistivity data (open circles) in the nano-Nb sample with $d_{XRD} \approx 5$nm. (b) Deviation from data of the fits to the single (dashed) and double (solid) exponentials. (c) Fit of the resistivity data (open circles) to double exponential equations (solid lines) for the three insulating samples. (Color online)

**Figure 7.** Dependence of grain boundary width ($\Delta d$, calculated from lattice constant data) on particle size ($D \equiv d_{XRD}$). Open circles indicate data points, while the error bars show the dispersion in particle size. The data can be fitted (solid line) to the empirical equation: $\Delta d = 0.48 + 1.86 \exp(-0.14D)$. Inset shows the variation of the higher barrier energy ($E_{g2}$) with the grain boundary width. The critical width of the grain boundary where the gap just opens up is indicated by $\Delta d_C$. (Color online)

**Figure 8.** Plots of ln $\sigma$ (conductivity) with (a) $T^{-1/4}$ and (b) $T^{-1/2}$ clearly show that the conductivity does not follow either relation over the entire temperature range of 4.2-300K.

**TABLE**

**Table 1.** Synthesis (dc sputtering) conditions and corresponding structural parameters for the nanostructured Nb films with different average crystallographic size ($d_{XRD}$). The spread in $d_{XRD}$ has been obtained from x-ray diffraction line shape analysis. (RT = room temperature)

| Dc power (W) | Gas pressure (mtorr) | Substrate temperature | Deposition time (h) | $d_{XRD}$ (nm) | Lattice parameter (nm) |
|---|---|---|---|---|---|
| 200 | 3 | 500°C | 3 | 60.4±10 | 0.3302 |
| 200 | 3 | 400°C | 3 | 43.0±8 | 0.3302 |
| 200 | 3 | 400°C | 2 | 28.2±5.2 | 0.3316 |
| 200 | 50 | RT | 4 | 19.2±3.1 | 0.3317 |
| 200 | 3 | RT | 1 | 18.2±3.0 | 0.3329 |
| 200 | 10 | RT | 1 | 17.6±3.0 | 0.3339 |
| 200 | 100 | RT | 3 | 10.7±1.8 | 0.3343 |
| 200 | 100 | RT | 4 | 8.2±1.4 | 0.3411 |
| 100 | 10 | RT | 1.5 | 7.0±1 | 0.3491 |
| 200 | 50 | RT | 1 | 6.0±1.1 | 0.3437 |
| 25 | 10 | RT | 2 | 5.0±1 | 0.3513 |

**Table 2**. Activation energies ($E_{g1}$ and $E_{g2}$) calculated from the transport data for nanostructured Nb samples with three different particle sizes ($d_{XRD}$). The corresponding grain boundary widths ($\Delta d$) have been calculated from x-ray diffraction data using a procedure explained in the text. The theoretical values of the Kubo gap ($\delta$) and the experimental value of the metal-insulator transition temperature ($T_i$) are also tabulated.

| $d_{XRD}$ (nm) | $\Delta d$ (nm) | $E_{g1}$ (meV) | $E_{g2}$ (meV) | $\delta$ (meV) calculated | $T_i = E_{g2}/k_B$ (K) | $\sigma_0$ $(\Omega m)^{-1}$ |
|---|---|---|---|---|---|---|
| 7 | 1.22 | 1.3 | 16 | 1.1 | 185 | 8342 |
| 6 | 1.28 | 2.5 | 25 | 1.7 | 292 | 7060 |
| 5 | 1.39 | 4.5 | 31 | 2.9 | 360 | 5900 |

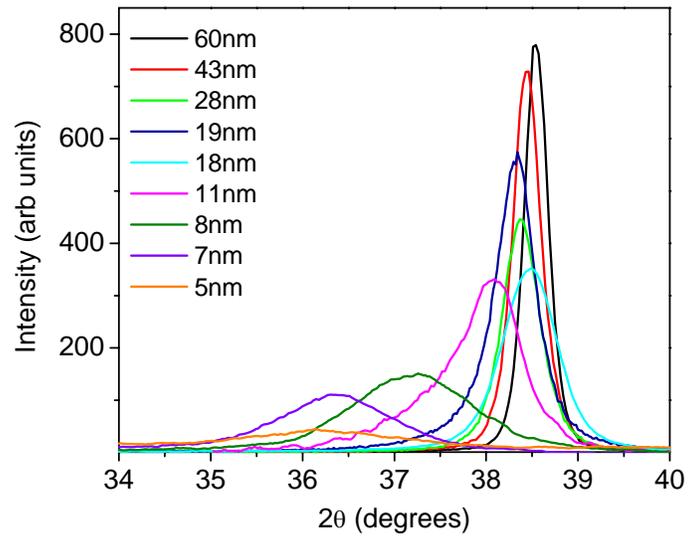

**Figure 1.** XRD pattern showing [110] reflection of *bcc* Nb for nanostructured Nb films with different values of coherently diffracting domain size ($d_{XRD}$).

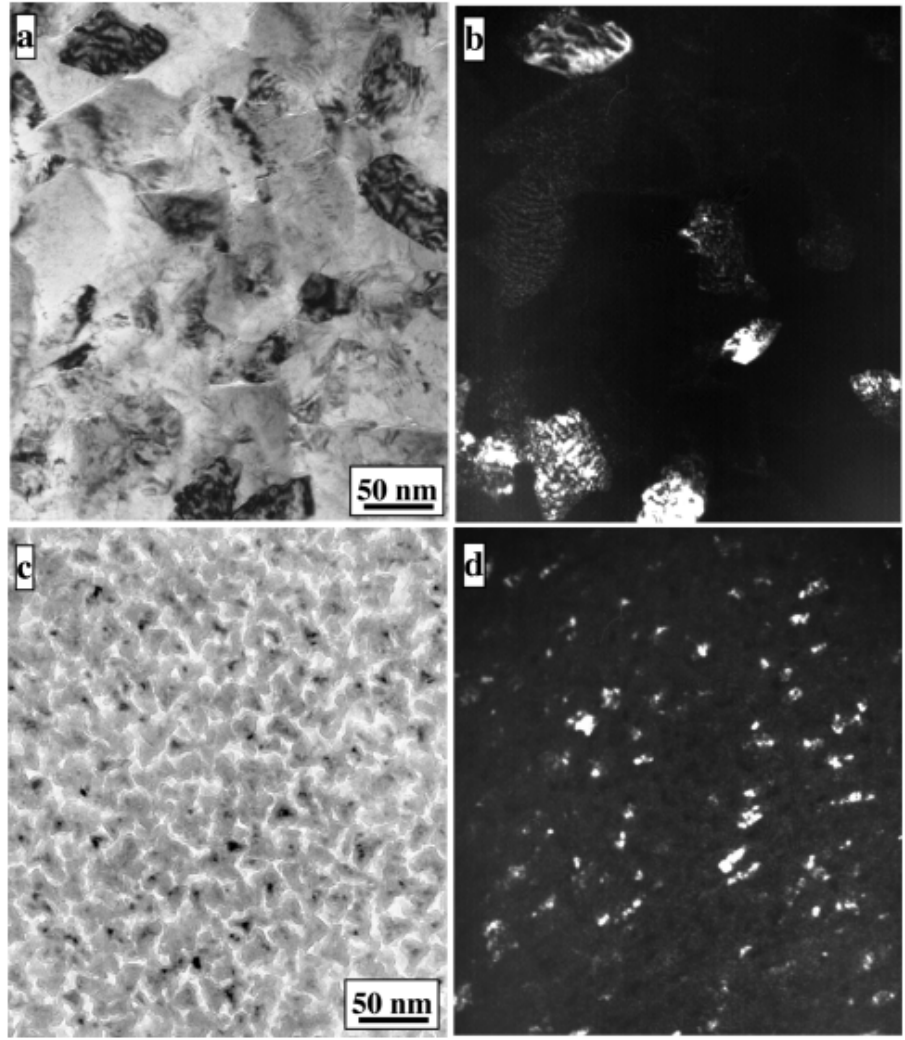

**Figure 2.** (a) Bright field and (b) dark field TEM images of the nanostructured Nb film with $d_{XRD} \approx 60$nm. (c) Bright field and (d) dark field TEM images of the film with $d_{XRD} \approx 5$ nm.

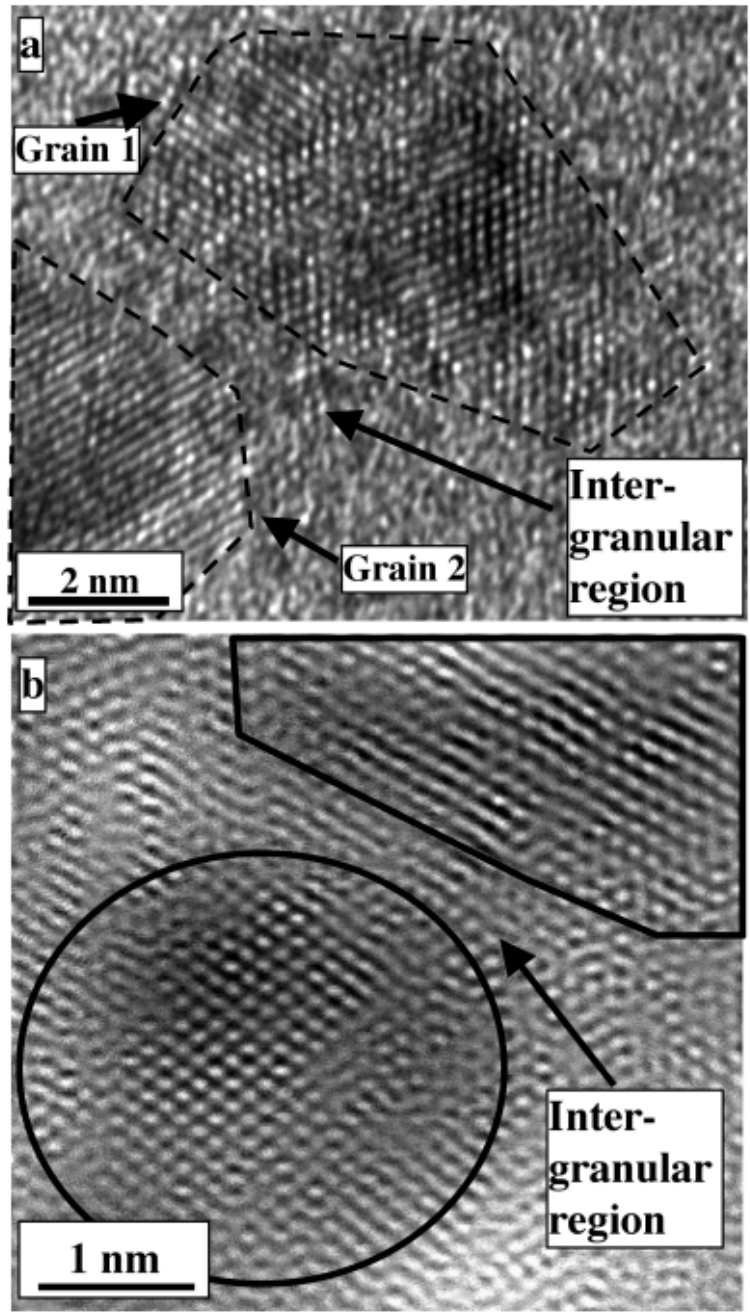

**Figure 3.** High-resolution transmission electron micrographs of the nanostructured Nb sample with $d_{XRD} \approx 5$nm. The grains are defined as crystallographically ordered regions (delineated in the figure) while the disordered inter-granular regions are $\approx 1$nm wide.

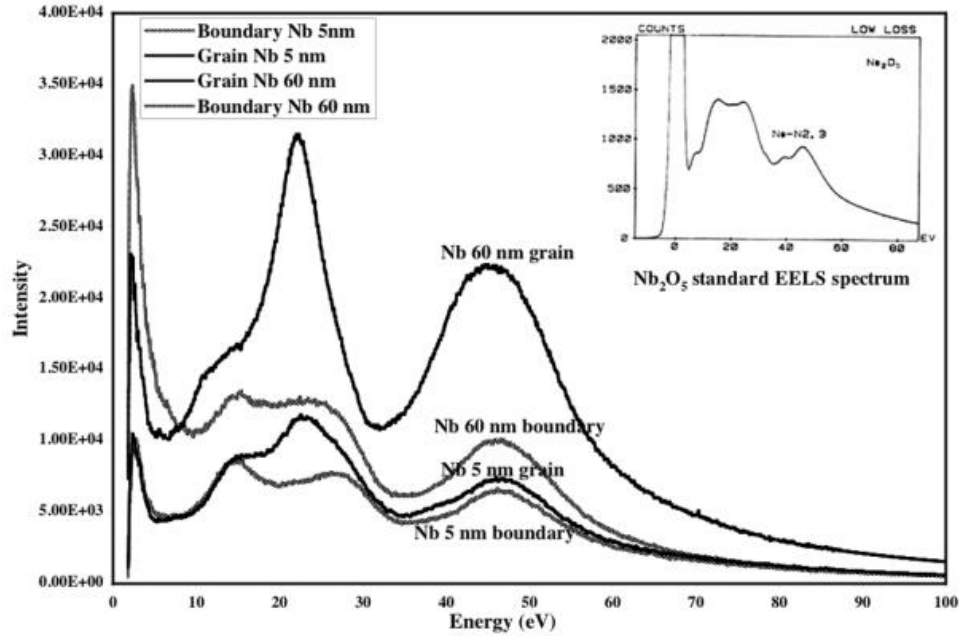

**Figure 4.** Electron energy loss spectra recorded (top to bottom, respectively) from the following specific regions i) interior of grains with $d_{XRD}$ = 18nm, ii) intergranular region of the 18nm sample, iii) interior of grains with $d_{XRD}$ = 5nm, and iv) intergranular region of the 5nm sample. A standard EELS spectrum from $Nb_2O_5$ is shown in the inset.

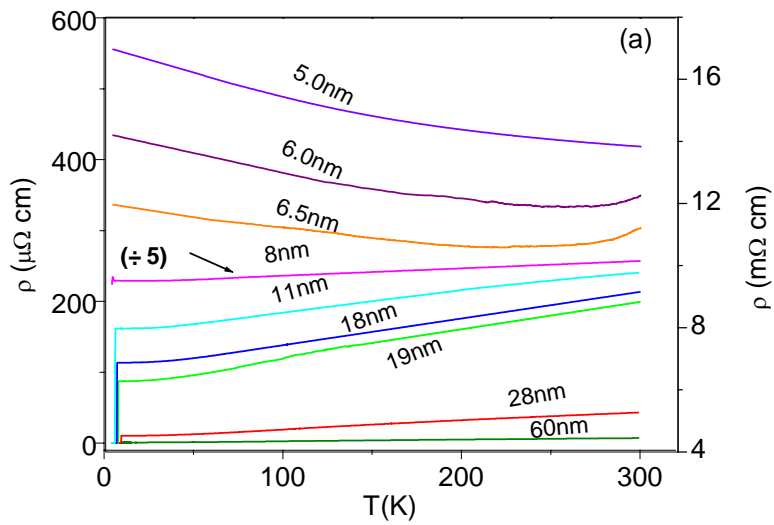

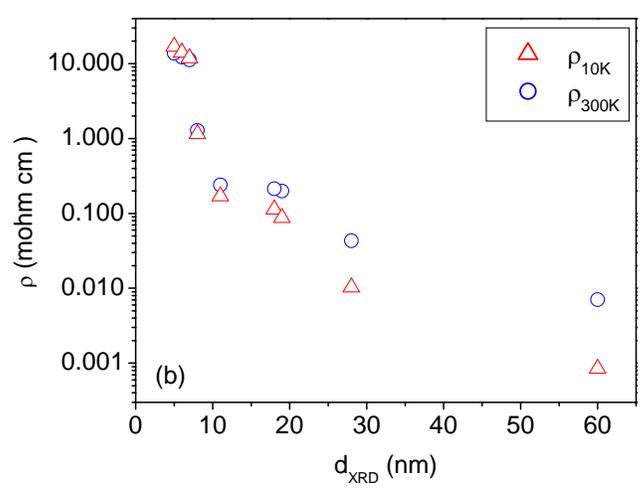

**Figure 5.** (a) Temperature dependence of the resistivity for nanostructured Nb films with different grain sizes ($d_{XRD}$). The scale on the left refers to the metallic films with $d_{XRD} \geq 8$nm while the scale to the right is for the insulating films with $d_{XRD} < 8$nm. (b) Particle size dependence of the resistivity at 10K and 300K, depicted, respectively by open triangles and circles.

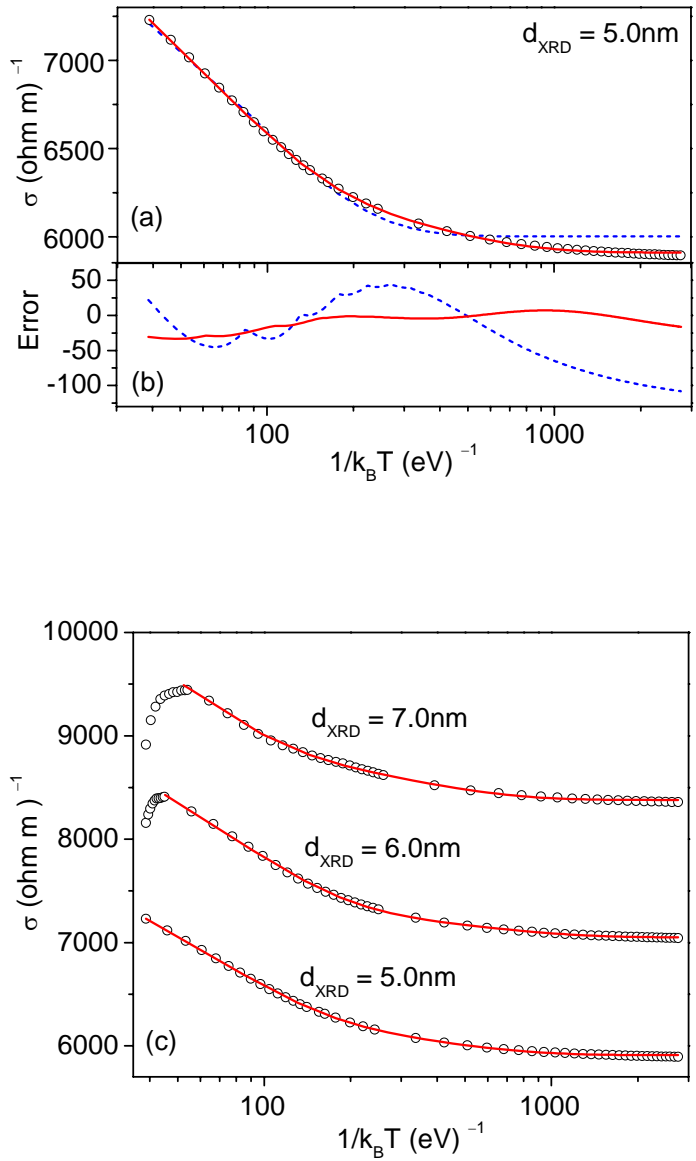

**Figure 6.** (a) Comparison of single (dashed line) and double (solid line) exponential fits to the resistivity data (open circles) in the nano-Nb sample with $d_{XRD} \approx 5$nm. (b) Deviation from data of the fits to the single (dashed) and double (solid) exponentials. (c) Fit of the resistivity data (open circles) to double exponential equations (solid lines) for the three insulating samples.

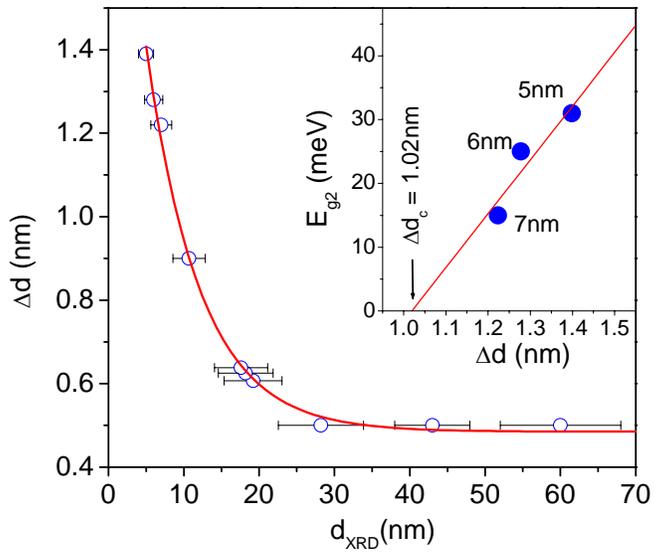

Figure 7. Dependence of grain boundary width (Δd, calculated from lattice constant data) on particle size ($D \equiv d_{XRD}$). Open circles indicate data points, while the error bars show the dispersion in particle size. The data can be fitted (solid line) to the empirical equation: $\Delta d = 0.48 + 1.86 \exp(-0.14D)$. Inset shows the variation of the higher barrier energy ($E_{g2}$) with the grain boundary width. The critical width of the grain boundary where the gap just opens up is indicated by $\Delta d_C$.

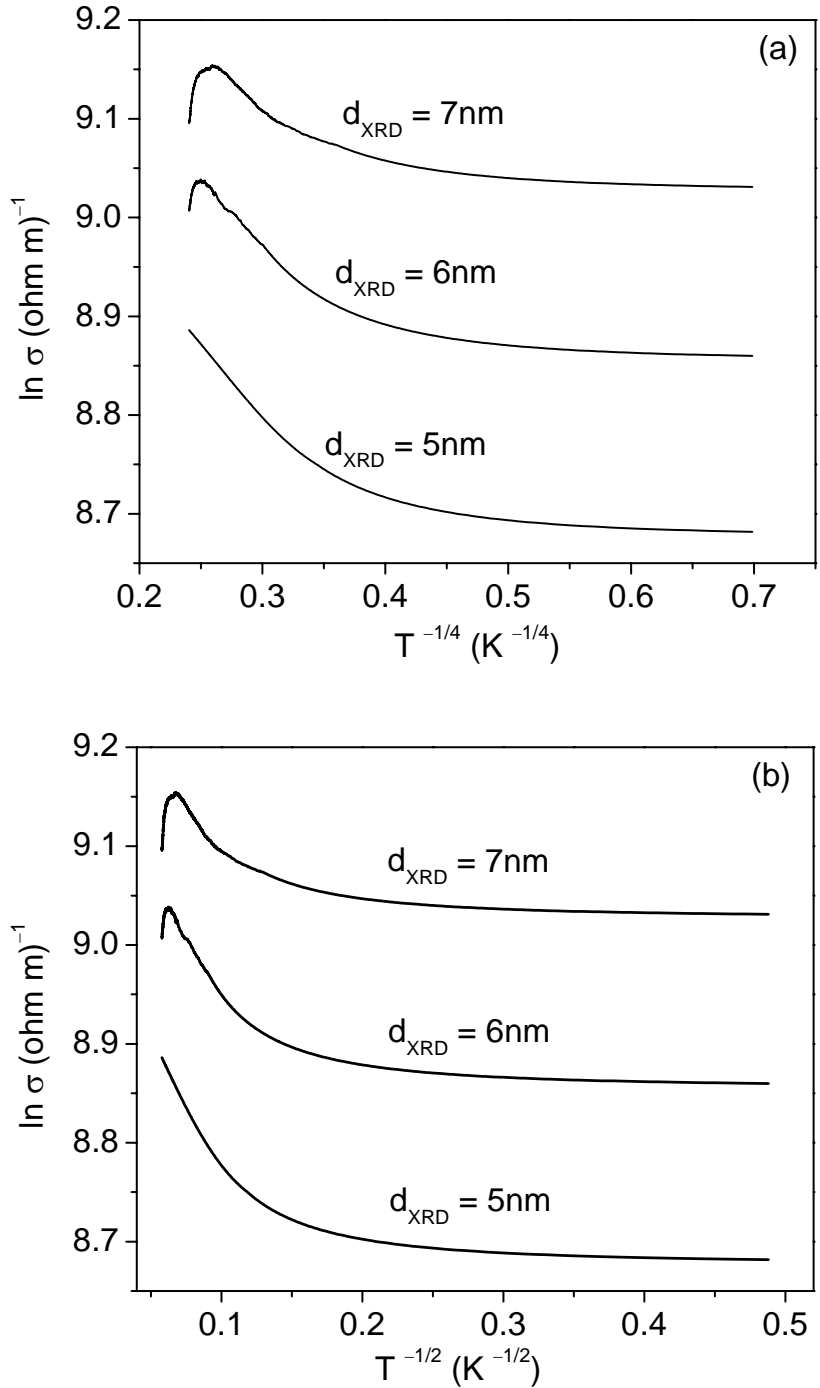

**Figure 8.** Plots of ln σ (conductivity) with (a) $T^{-1/4}$ and (b) $T^{-1/2}$ clearly show that the conductivity does not follow either relation over the entire temperature range of 4.2-300K.